\documentclass[10pt, twocolumn]{aastex63} 

\usepackage[english]{babel}
\usepackage[utf8x]{inputenc}

\usepackage{natbib}
\usepackage{graphicx}
\usepackage{hyperref}

\newcommand{\accuracy}{93.5\%}
\newcommand{\precision}{97.7\%}
\newcommand{\recall}{92.4\%}
\newcommand{\classifier}{\texttt{GWSkyNet}}

\begin{document}

\title{\classifier: a real-time classifier for public gravitational-wave candidates}

\correspondingauthor{Miriam Cabero} 
\email{micamu@phas.ubc.ca}

\author[0000-0003-4059-4512]{Miriam Cabero}
\affiliation{Department of Physics and Astronomy, The University of British Columbia, Vancouver, BC V6T 1Z4, Canada}

\author[0000-0003-2242-0244]{Ashish Mahabal}
\affiliation{Division of Physics, Mathematics, and Astronomy, California Institute of Technology, Pasadena, CA 91125, USA}
\affiliation{Center for Data Driven Discovery, California Institute of Technology, Pasadena, CA 91125, USA}

\author[0000-0003-0316-1355]{Jess McIver}
\affiliation{Department of Physics and Astronomy, The University of British Columbia, Vancouver, BC V6T 1Z4, Canada}

\begin{abstract}
The rapid release of accurate sky localization for gravitational-wave candidates is crucial for multi-messenger observations. During the third observing run of Advanced LIGO and Advanced Virgo, automated gravitational-wave alerts were publicly released within minutes of detection. Subsequent inspection and analysis resulted in the eventual retraction of a fraction of the candidates. Updates could be delayed by up to several days, sometimes issued during or after exhaustive multi-messenger followup campaigns. We introduce \classifier, a real-time framework to distinguish between astrophysical events and instrumental artefacts using only publicly available information from the LIGO-Virgo open public alerts. This framework consists of a non-sequential convolutional neural network involving sky maps and metadata. \classifier{} achieves a prediction accuracy of \accuracy{} on a testing data set. 
\end{abstract}


\section{Introduction}
\label{sec:introduction}

The LIGO Scientific and Virgo Collaborations have so far reported fifteen confirmed gravitational-wave (GW) signals from the mergers of compact objects \citep{GWTC-1, GW190425, GW190412, GW190814, GW190521}. Depending on the nature of the objects and the environment surrounding the binary, merging compact binaries can also emit electromagnetic radiation and neutrinos. In the second observing run (O2) of Advanced LIGO and Advanced Virgo, low latency GW candidates were distributed to partner astronomers for multi-messenger followup \citep{LIGOScientific:2019gag}. This partnership resulted in the combined measurement of electromagnetic and GW emission from GW170817 \citep{GW170817, GBM:2017lvd}, which revolutionized our understanding of a wide range of physics related to merging neutron stars. 

Accurate sky localization and the rapid release of GW candidates are crucial for multi-messenger observations of astrophysical events. The former is achieved through a global network of GW observatories: the two Advanced LIGO detectors in the United States (Hanford and Livingston)  together with the Advanced Virgo detector in Italy have been able to pinpoint the sky location of nearby GW candidates to areas smaller than 50 square degrees. The latter is achieved through fast and efficient data analysis pipelines. During the third observing run (O3) of Advanced LIGO and Advanced Virgo, low latency GW alerts were publicly available within minutes through the Gamma-ray Coordinates Network (GCN)\footnote{\url{https://gcn.gsfc.nasa.gov/}} and on the GW candidate event DataBase (GraceDB)\footnote{\url{https://gracedb.ligo.org/}}. For compact binaries, the information released in Open Public Alerts (OPA) included\footnote{\url{https://emfollow.docs.ligo.org/userguide/index.html}}: the false-alarm-rate (FAR) estimated by the search pipelines, the inferred sky position and distance of the source \citep{Singer:2015ema, Singer:2016eax}, the probability of astrophysical origin with classification according to different mass regions \citep{Kapadia:2019uut}, and the probability of neutron star matter content \citep{Chatterjee:2019avs, Foucart:2018rjc}. Rapid and public information release enabled every astronomer with access to telescopes or neutrino detectors to search for counterparts to the GW candidates \citep{Andreoni:2019kqi, Antier:2020nuy, Hussain:2019xzb, Vieira:2020lze, Kasliwal:2020wmy, Graham:2020gwr, Coughlin:2020pbb, Coughlin:2020fwx}. 

Preliminary OPAs in O3 were generated automatically when low-latency GW search pipelines reported an effective FAR below either 1 per 2 months for the matched-filtering searches or 1 per year for the unmodeled burst searches \citep{Aasi:2013wya}. This FAR threshold targets an overall astrophysical purity of 90\%, hence some candidates will likely originate from instrumental artefacts. GW candidates recognized as the result of an instrumental artefact after human inspection were retracted by LIGO and Virgo. Retractions usually appeared within an hour of the initial alert, although in some occasions it could take significantly longer, up to several days. Some surviving candidates were updated when further analyses provided more accurate information, sometimes changing the binary classification. LIGO and Virgo released a total of 80 OPAs during O3, of which 24 were subsequently retracted. Of the remaining 56 candidates, 26 were reported with a non-vanishing probability of being of terrestrial origin, and 4 have been confirmed so far as real astrophysical events by the LIGO Scientific and Virgo Collaborations \citep{GW190425, GW190412, GW190814, GW190521}. 

Telescope time is competitive and astronomers have to carefully select which candidates to pursue. Some of the retracted events in O3 had very accurate sky localization or very low FARs, qualities that make a candidate promising for followup. For instance, the binary neutron star candidate S190822c\footnote{\url{https://gracedb.ligo.org/superevents/S190822c/view/}} was reported with a FAR of 1 per 5x$10^9$ years. Within less than one hour, three different followups with various instruments were reported \citep{S190822c_followup_Cook, S190822c_followup_Lipunov, S190822c_followup_Hussain}. The candidate was then retracted 1.5 hours after the preliminary alert. Depending on the time of the event and the available instruments, telescopes can be steered to point towards the sky location of GW candidates within minutes (though counterpart identification, if achieved, can take a much longer time depending on the area of uncertainty of the event). How can astronomers best decide which candidates to follow up based on the information released in OPAs? 

In this work we introduce \classifier, a low latency classifier to separate real astrophysical events from terrestrial artefacts, supplementing information provided by the LIGO Scientific and Virgo Collaborations. Contrary to recent GW classifiers \citep{Chatterjee:2019avs, Kim:2019ktw}, which rely on access to LIGO-Virgo data, \classifier{} only requires publicly available data products and hence can be used by any astronomer. Built with state-of-the-art machine learning algorithms such as non-sequential convolutional neural networks and multiple input machine learning models, \classifier{}  achieves an accuracy of \accuracy{} on a testing data set.

This manuscript is organized as follows. In Sec.~\ref{sec:dataset} we describe the data set constructed to train and test the machine learning algorithm. In Sec.~\ref{sec:CNN} we describe the data preparation and the features selected for training, as well as the architecture chosen for the machine learning model. Section~\ref{sec:results} presents the results obtained with the fully trained model, including predictions for the O3 candidates in GraceDB. Finally, we summarize our findings in Sec.~\ref{sec:conclusion}.

\section{The data set} \label{sec:dataset}

A data set with a large, balanced number of labeled examples is key to the success of supervised machine learning. With GW astronomy still in its infancy, the currently confirmed population of astrophysical GW events is insufficient to construct a robust training set. This section describes the methods used to simulate a population of real candidates and identify a population of noise candidates in real data. 

\subsection{Real events}

A population of simulated real events is constructed using gravitational waveform models for compact binaries, available in the \texttt{LALSuite} \citep{lalsuite} package. Masses are drawn from a uniform distribution in the total mass of the binary, with component masses in the range $m \in [1.2, 2.2] M_\odot$ for neutron stars and $m \in [3, 100] M_\odot$ for black holes. For simplicity, spins are restricted to be along the direction of orbital angular momentum. Spin magnitudes are constrained to $\chi \leq 0.05$ for neutron stars and $\chi \leq 0.99$ for black holes. With these conditions, we generate three types of binaries: binary neutron stars, neutron-star black-hole binaries and binary black holes. Sources are distributed uniformly in co-moving volume with a maximum distance of up to 1.5 Gpc, depending on the binary type and the detectors sensitivities. Sky locations are randomly distributed, and the inclination angle, $\iota$, is distributed uniformly over $\arccos{\iota}$.

To avoid contaminating the training set of real events with noise artefacts present in GW data, simulated signals are added into Gaussian noise coloured with the power spectral densities (PSD) of Advanced LIGO and Advanced Virgo. Three PSDs associated with three different GW events in the Gravitational-Wave Open Science Center (GWOSC) \citep{Abbott:2019ebz} are chosen to represent different sensitivities of the detectors.
The PSDs at the time of GW150914 and GW170104 represent the two Advanced LIGO detectors (Hanford and Livingston) in the first (O1) and second (O2) observing runs, respectively. To incorporate the Advanced Virgo detector, we additionally use the PSD at the time of GW170729, when Virgo had joined the observing run and the two LIGO detectors had significantly different sensitivities with respect to the rest of O2 \citep{GWTC-1}. 
Simulated signals are distributed equally among these three different representative sensitivities.

To select only those signals that would be detectable by GW search pipelines, we impose the following conditions on the matched-filtering signal-to-noise ratio (SNR) of the signal. First, the signal must be detected in at least two observatories with SNR $\rho \geq 4.5$. Second, the network SNR, given by the quadrature sum of the individual SNRs satisfying the first condition, must be $\rho_N \geq 7$. After applying these conditions, the final data set of simulated real events consists of 2857 candidates. While this population of real events can be easily increased, it is important to keep a good balance in the training set between real and noise candidates \cite{johnson2019survey}. 

\subsection{Noise events}

\begin{figure}[t]
    \centering
    \includegraphics[width=\columnwidth]{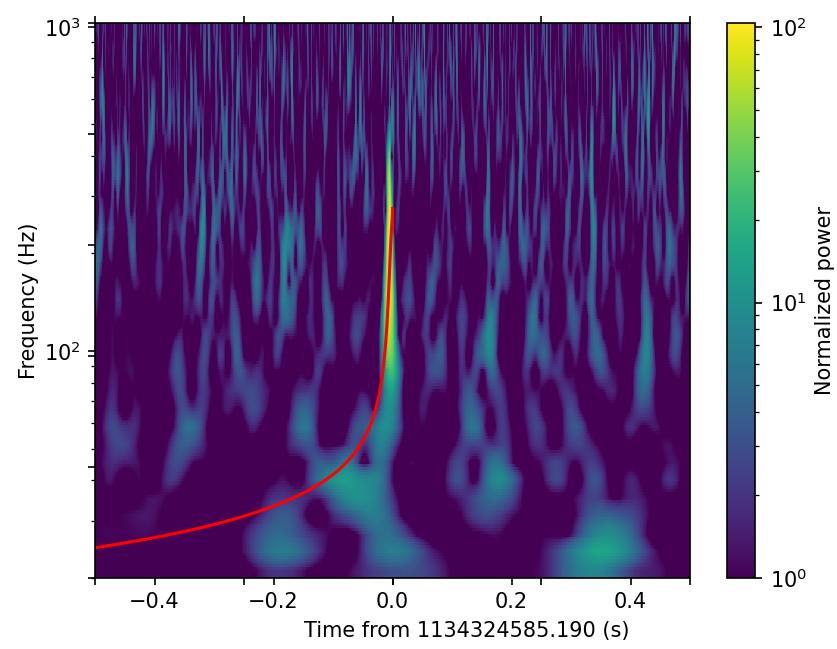}
    \caption{Time-frequency representation of a blip glitch included in the noise set. The red curve represents the GW template that triggered the 2-OGC search. This template corresponds to a binary black hole with masses $m_1 = 40 M_\odot$ and $m_2 = 19 M_\odot$. The network SNR reported by the search at the time of this glitch is $\rho_N = 11.3$.}
    \label{fig:glitch}
\end{figure}

Matched-filtering searches for GWs from compact binaries select candidates by requiring a temporal coincidence between detectors \citep{Nitz:2018rgo, Messick:2016aqy, Adams:2015ulm, Chu2017}. The coincidence window is set by the maximum light-travel time between detectors plus the uncertainty in the inferred coalescence time in each detector. Noise transients (glitches) are independent between detectors and will rarely happen simultaneously. However, random noise fluctuations can have a sufficiently high SNR to trigger a coincidence in the search pipelines at the time of a glitch. If the resulting network SNR is greater than the pipeline's predetermined threshold (and the FAR is sufficiently low), the glitch can be identified as a GW candidate by the low-latency searches and generate an automatic OPA.

Two main ingredients are required to identify a population of noise events: a collection of known glitches in real detector data, and a collection of corresponding candidates (triggers) from the GW search pipelines. For the former, we use the collection of blip glitches from \cite{Cabero:2019orq} and the O1 and O2 collections of glitches from the Gravity Spy classifier \citep{Zevin:2016qwy, bahaadini2018machine, Coughlin:2019ref}. For the latter, we use the search triggers from the second Open Gravitational-wave Catalog (2-OGC) \citep{Nitz:2019hdf}. Each of these catalogs contain thousands of glitches and triggers, but only a small fraction of glitches will be associated with a coincident trigger and satisfy the SNR requirements imposed by the search pipelines. 

While the temporal coincidence condition is consistent among GW searches, different search pipelines use different SNR requirements. Here, we apply the same threshold as for the simulated signals of the real events set: individual SNR $\rho \geq 4.5$ in at least two detectors, and network SNR $\rho_N \geq 7$. When a glitch is within $\pm 0.5$s of a 2-OGC coincident search trigger in the same detector, the candidate is selected for further evaluation. All the selected glitches are then visually inspected to ensure that the glitch and the waveform overlap. Figure \ref{fig:glitch} shows an example of a noise candidate, with the gravitational waveform of the coincident trigger overlaid on top of the glitch. The final data set of noise events consists of 1267 candidates, which makes 30\% of the total training set. In the future, with the release of the O3 data sets, additional glitches will become available for the training set.

\section{Machine learning classifier} \label{sec:CNN}

Convolutional Neural Networks (CNN) are state-of-the-art deep learning algorithms commonly used for image classification. We employ a non-sequential CNN architecture to design \classifier, a real/noise classifier for real-time GW-candidate alerts. The main inputs to the classifier are the 3D sky map images released in the LIGO-Virgo OPAs. During O2, a method based on the mutual information distance of 2D GW sky maps was included in part of the LIGO-Virgo human vetting to distinguish between signal and noise \citep{Essick2017, Essick_github}. Here we introduce the first CNN classifier for GW sky maps, with a more diverse set of inputs. We find that additional numerical and categorical data, such as the distance to the source and the detector network, are important for sky map classification. Therefore, we construct a multiple input CNN model that accepts mixed data features. In this section, we first explain the features chosen for training and then describe the specifications of the network architecture.

\subsection{Training features}

\begin{figure}[t]
    \centering
    \includegraphics[width=\columnwidth]{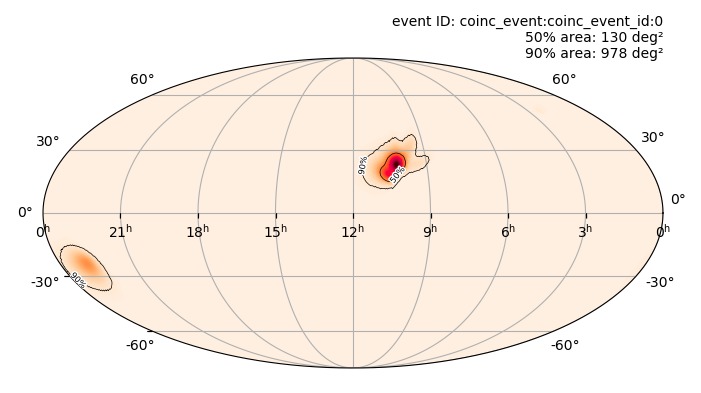}
    \includegraphics[width=\columnwidth]{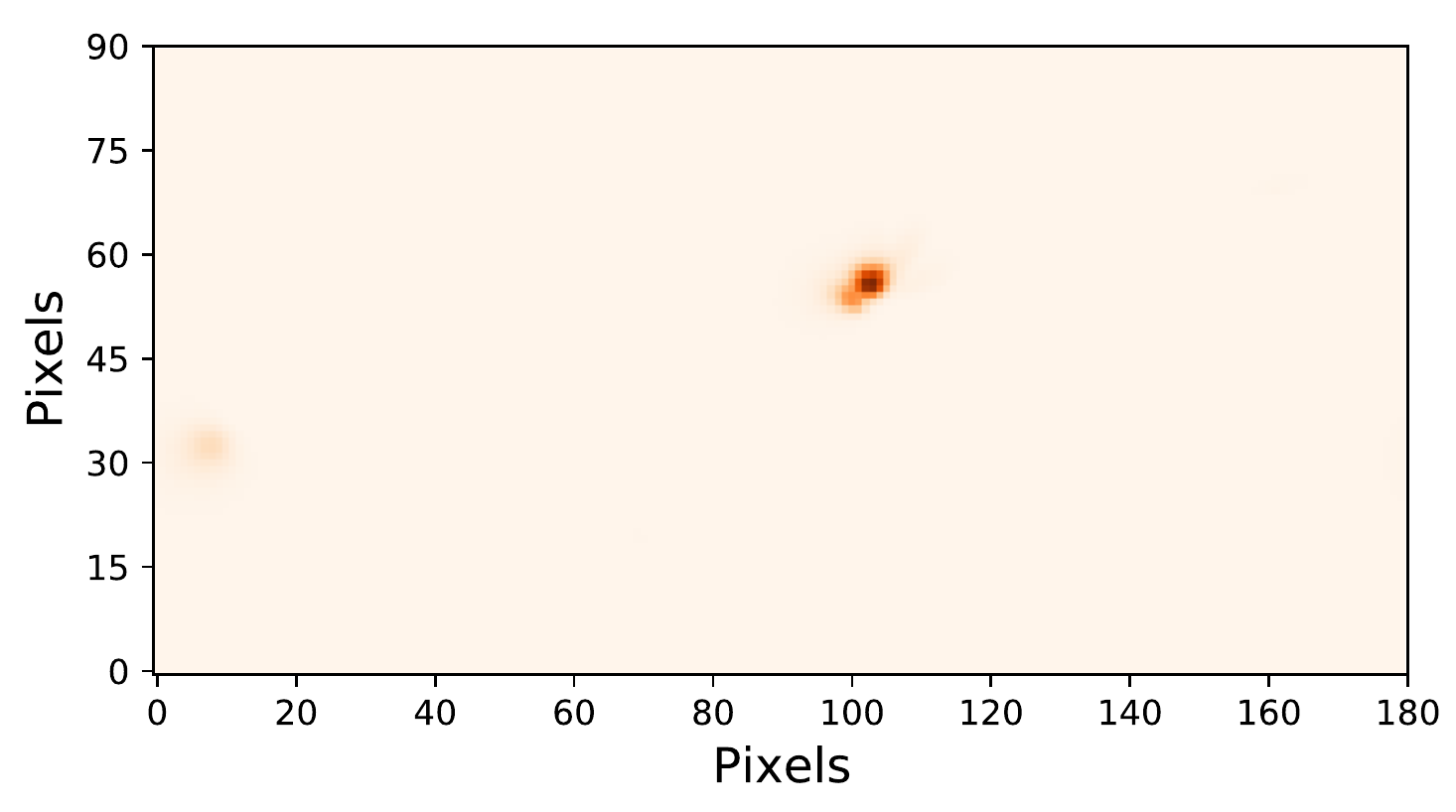}
    \caption{Sky map obtained with \texttt{BAYESTAR} for the blip glitch shown in Fig. \ref{fig:glitch}. (Top) Image as it would appear in an OPA on GraceDB. (Bottom) Same image after re-projection, normalization and down-sampling onto a 90x180 grid. This image is used for training the machine learning algorithm.}
    \label{fig:glitch_skymap}
\end{figure}

We use \texttt{BAYESTAR} \citep{Singer:2015ema, Singer:2016eax, Singer:2016erz} to reconstruct the 3D sky localization of the source for each candidate in the training set. The sky localization is stored as a Flexible Image Transport System (FITS) file. \texttt{BAYESTAR} FITS files contain four columns that represent four channels of a Hierarchical Equal Area isoLatitude Pixelization (HEALPIx) all-sky image. The first column (\textsc{prob}) represents the 2D probability sky map. The other three columns (\textsc{distmu}, \textsc{diststd} and \textsc{distnorm}) are used to calculate the 3D probability volume map. The \texttt{BAYESTAR} FITS header provides additional metadata about the candidate. Of particular interest for the purpose of this work are the posterior mean distance of the source and the list of GW instruments that triggered the candidate.

For training, we transform each column in the FITS file (the sky map images) to a rectangular projection with 180\,\textrm{x}\,360 pixels. Pixels with invalid values in the volume columns, as defined in \cite{Singer:2016erz}, or with NaN values originating from the re-projection, are replaced with zero. Each image is then normalized independently dividing by its maximum value. To reduce computations during training, images are down-sampled to 90\,\textrm{x}\,180 pixels using the maximum pooling method. Figure \ref{fig:glitch_skymap} shows an example of a sky map image after re-projection, normalization and down-sampling. The three volume images are stacked together to form a 3-channel cuboid that will be used as input for the classifier. The final set used as input for training the machine learning algorithm consists of eight features:
\begin{itemize}
    \item Sky map image, shape (90, 180, 1),
    \item Stacked volume images, shape (90, 180, 3),
    \item Posterior mean distance in Mpc, normalized dividing by the maximum value in the training set,
    \item Detector network in 3-bit multi-hot encoding format, with one bit reserved for each detector (LIGO Hanford, LIGO Livingston and Virgo),
    \item Four normalization factors (one for each image), normalized dividing by the corresponding maximum value in the training set.
\end{itemize}

\subsection{\classifier{} architecture and training}

\begin{figure}[t]
    \centering
    \includegraphics[width=\columnwidth]{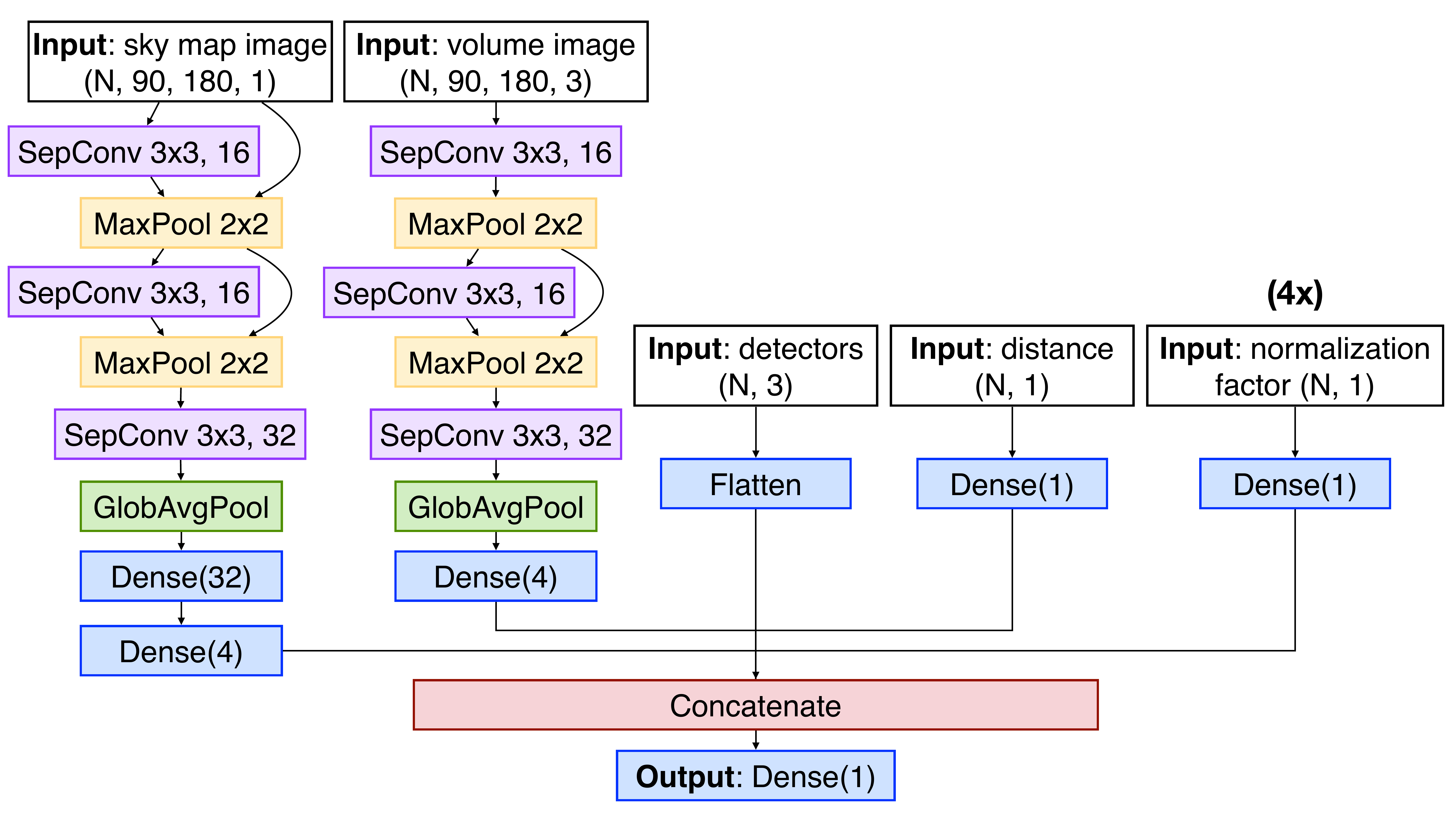}
    \caption{Architecture of the multiple input model constructed to classify GW sky maps. The first two branches are CNNs with residual connections for image data. The shape of the input data is indicated in parenthesis, with N the number of examples in the training set. The numbers in the SeparableConv2D (SepConv) and MaxPool layers indicate the kernel size in pixels and the number of filters (when applicable). The number in the Dense layers indicates the number of units.}
    \label{fig:architecture}
\end{figure}

We use \texttt{TensorFlow} \citep{tensorflow2015-whitepaper} and the functional API in \texttt{Keras} to construct a multiple input model with eight branches. The first two branches are CNNs designed to operate over the image data (sky map and 3-channel volume images). To improve the performance of the classifier, residual connections are added to each CNN branch, with two residual blocks in the sky map branch and one block in the volume branch. The other six branches are simple multi-layer perceptrons for the categorical and numerical data. All branches are concatenated together into the final multiple input model. A complete description of the specifications of the network can be seen in Fig. \ref{fig:architecture}. Rectified Linear Units (ReLU) activation functions are used for all hidden layers, and a sigmoid activation function is used for the output layer.

The training set introduced in Sec. \ref{sec:dataset}, with 4124 candidates, is split into 81\% for training, 9\% for validation and 10\% for testing. The algorithm is trained on GPUs available through Google Colaboratory\footnote{\url{https://colab.research.google.com/notebooks/intro.ipynb}}, and hyperparameter optimization is performed with \texttt{KerasTuner} \citep{omalley2019kerastuner}. Training is enabled for 500 epochs, employing the binary cross-entropy loss function, the Adam optimizer with a learning rate of 0.002, and batch size of 32. Our use of GlobalAveragePooling ensured a small model (3687 free parameters in total), helping us avoid overfitting.

\section{Performance and predictions} \label{sec:results}

\begin{figure}[t]
    \centering
    \includegraphics[width=\columnwidth]{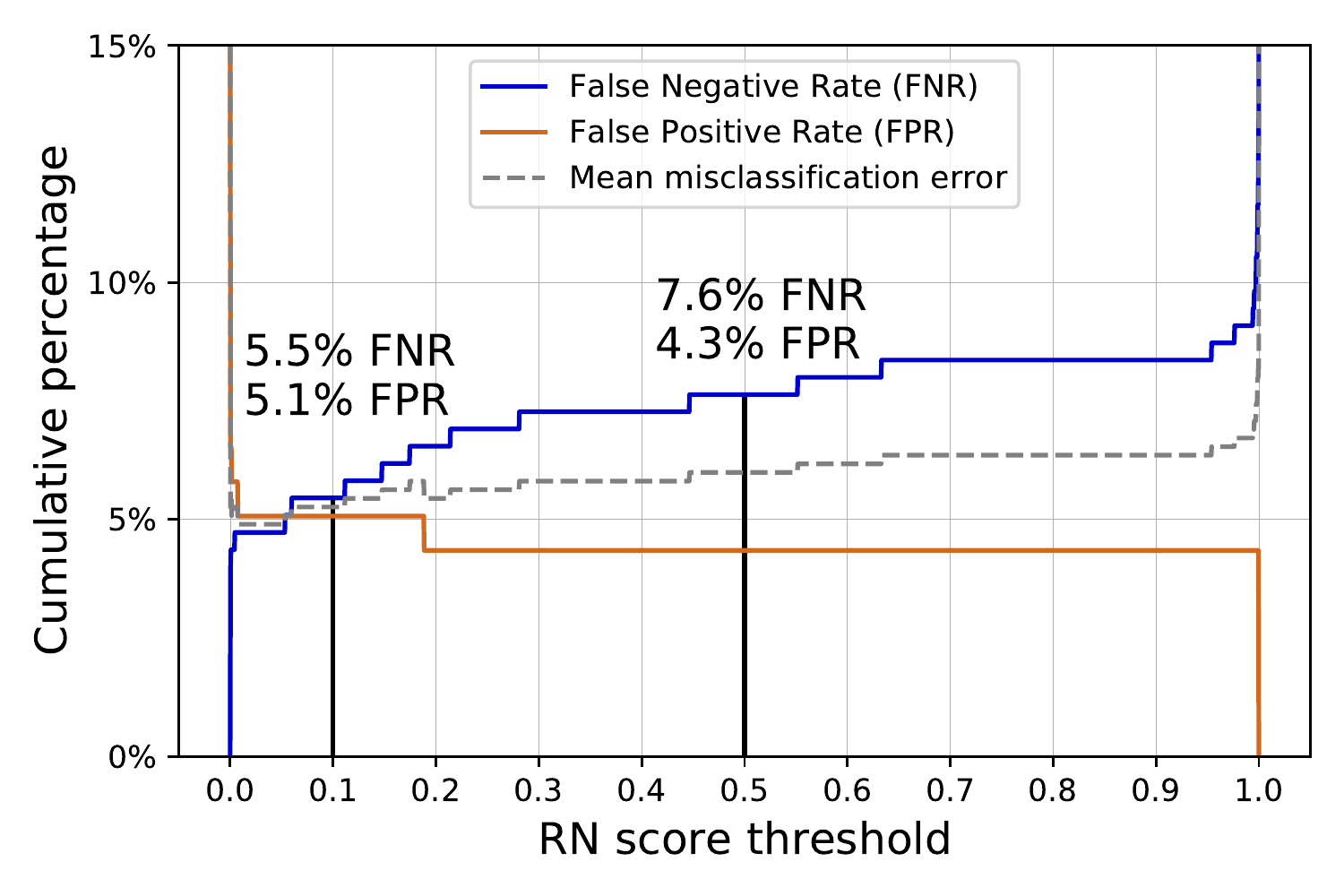}
    \caption{False Positive Rate (FPR) and False Negative Rate (FNR) as functions of the Real-Noise (RN) score threshold (based on Figure 9 in \cite{duev2019real}). At a score threshold RN$\geq0.5$, \classifier{} yields 7.6\% FNR and 4.3\% FPR. Lowering the threshold to RN$\geq0.1$ reduces the FNR to 5.5\%, with small variation on the FPR (5.1\%). The intersection is at RN$\simeq0.06$, with FPR$=$FNR$=$5.1\%.}
    \label{fig:performance}
\end{figure}

\begin{figure}[hbt!]
    \centering
    \includegraphics[width=0.7\columnwidth]{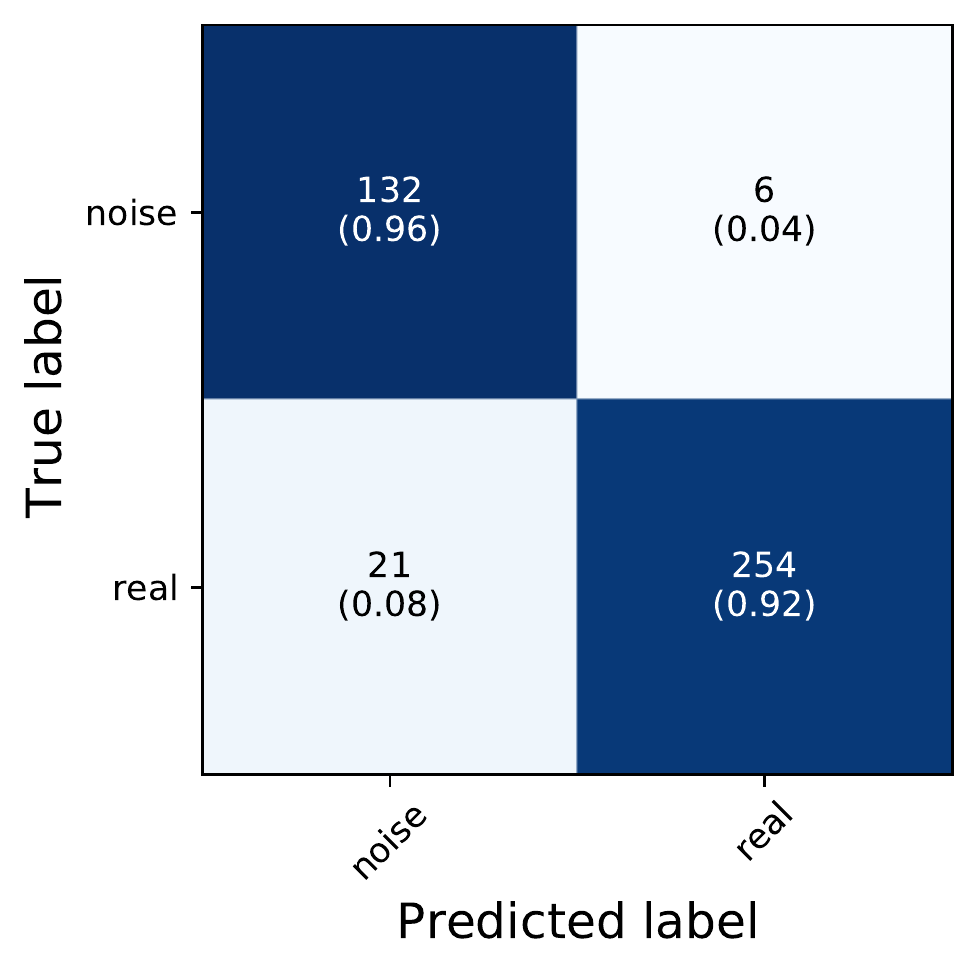}
    \includegraphics[width=0.7\columnwidth]{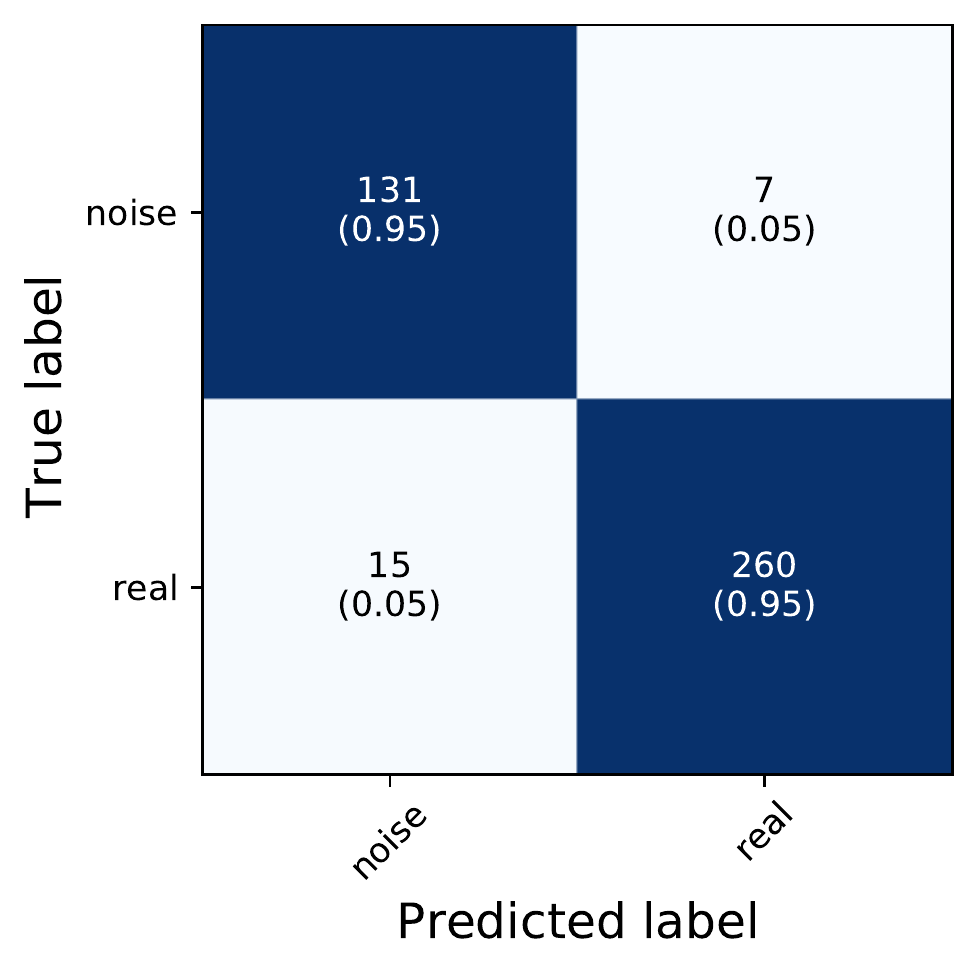}
    \caption{Confusion matrix, with normalized values in parenthesis, for thresholds RN$\geq 0.5$ (top) and RN$\geq 0.1$ (bottom) denoting a real source.}
    \label{fig:confusion}
\end{figure}

The accuracy, which measures the fraction of correct predictions over the total number of classifications, is a performance metric commonly used for binary classifiers. With a Real-Noise (RN) score threshold of RN$\geq 0.5$ denoting a real source, the fully trained \classifier{} achieves an accuracy of \accuracy{} on a test set with 413 candidates, independent from the training set. We also use precision and recall values as additional performance measures. The precision, defined as the fraction of true predicted positives over the total number of predicted positives, measures how often the model is correct when it predicts that a candidate is real. The recall or true-positive rate, defined as the fraction of true predicted positives over the total number of actual positives, measures how many real candidates are predicted correctly. On the test set, the classifier achieves a precision of \precision{} and a recall of \recall{} for RN$\geq 0.5$. Figure \ref{fig:performance} shows the False Positive Rate (FPR) and False Negative Rate (FNR) on the test set for different Real-Noise (RN) score thresholds. The FPR varies very slowly with the RN threshold, so one can lower the FNR by lowering the threshold with only a small increase in the FPR. In Fig. \ref{fig:confusion} we show the confusion matrices on the test set for two different RN thresholds of 0.1 and 0.5. Based on one's preference for higher recall or higher precision, the threshold can be changed with the understanding that improving recall (precision) involves lowering precision (recall).

The O3 retracted candidates and confirmed astrophysical signals are a further test of the classifier performance. For each candidate, we use the \texttt{BAYESTAR} multi-order FITS files from GraceDB, since the training set was built using this pipeline. Of the 4 confirmed GW signals \citep{GW190425, GW190412, GW190814, GW190521}, the classifier recognizes all of them as true astrophysical signals. Of the 22 retracted candidates identified by matched-filtering searches, 10 are correctly identified as instrumental artefacts and 12 are misclassified as astrophysical. Among the correct classifications is, for instance, the candidate S190822c mentioned in Sec. \ref{sec:introduction}, which was retracted 1.5 hours after the preliminary alert. In future observing runs, \classifier{} could provide immediate warning about the plausibility of a candidate, in some occasions sparing followup time invested on non-astrophysical candidates. The misclassifications on noise candidates can be explained from the data set used for training. On one hand, most of the retracted candidates were identified by the \texttt{GstLal} and \texttt{MBTAOnline} search pipelines (as reported in GraceDB), while all the noise candidates in the training set originated from \texttt{PyCBC} search triggers. On the other hand, almost 70\% of the retracted candidates included the Virgo detector, for which we could not include many noise events in the training set due to the small amount of available Virgo data. Incorporation of additional glitches in the training set, both by including results from different search pipelines as well as glitches from O3 data (when available), should lead to improved results. In addition, visualization of layers of the trained network can provide an insight into the decisions reached. This is a nascent field and we will be exploring related methodology for building in more interpretability and explainability.

\subsection{O3 GraceDB predictions}

We report predictions for the 51 unpublished candidates identified by at least one matched-filtering search for compact binaries. Using the initial \texttt{BAYESTAR} multi-order FITS files released in the preliminary alert, \classifier{} predicts that 41 candidates are of astrophysical origin, and 10 candidates are due to terrestrial artefacts: S190510g, S190718y, S190901ap, S190910d, S190910h, S190930t, S191105e, S191205ah, S200213t and S200302c. All these 10 candidates have a non-zero probability of terrestrial origin, as estimated by LIGO and Virgo methods \citep{Kapadia:2019uut}. However, two of these candidates are single-detector triggers (S190910h and S190930t). Since there are no examples in the training set of GW candidates observed by only one detector, predictions on single-detector candidates should be taken with caution. The release of the O3 catalog by the LIGO Scientific and Virgo Collaborations will determine the accuracy of these predictions. 

\section{Conclusion} \label{sec:conclusion}

In this work we have introduced \classifier, a new machine learning binary classifier for low-latency GW candidates that complements the information released by the LIGO Scientific and Virgo Collaborations. This classifier, based solely on publicly available data products, can be used by the broader astronomy community for fast decision making on candidate followup. In future observing runs, real time predictions from \classifier{} on preliminary automated alerts could identify noise candidates without the delay of human-based retractions or analysis updates. More importantly, astronomers will receive additional information on the astrophysical probability of non-retracted candidates.

The classifier has been trained using LIGO and Virgo detector glitches as noise candidates and simulated signals of compact binary mergers as real candidates. With a threshold $RN\geq0.5$ denoting a real source., the accuracy achieved on a test data set, which is independent of the training set, is \accuracy{}. Additionally, we reported predictions from \classifier{} on O3 GraceDB candidates. Of the 51 unpublished non-retracted candidates, the classifier identified 10 as not astrophysical, with the caveat that two of those are single-detector candidates.  

Increasing the size of the training set will be crucial to improve the classifier performance. In future work, we will incorporate additional candidates from different search pipelines, as well as from the O3 data set. With a larger training set in hand, we will construct a multi-class classifier capable of distinguishing between different classes of binary mergers.
We provide the classifier and its data sources on GitHub\footnote{https://github.com/GWML/GWSkyNet} and preserve the current version at the CaltechDATA repository: doi:10.22002/D1.1659.

\acknowledgments
{We are thankful to the Gravity Spy team, and in particular Scott Coughlin, for sharing the collection of O1 and O2 glitches. We are also thankful to Deep Chatterjee, Tito Dal Canton, Derek Davis, Daryl Haggard, Ian Harry, Fergus Hayes, Alexander H. Nitz, Leo P. Singer, Nicholas Vieira and the anonymous referee for useful comments and discussions. 
MC and JM acknowledge funding from the Natural Sciences and Engineering Research Council of Canada (NSERC). AM acknowledges support from the NSF (1640818, AST-1815034). AM and JM also acknowledge support from IUSSTF (JC-001/2017).
This research has made use of data, software and/or web tools obtained from the Gravitational Wave Open Science Center (\url{https://www.gw-openscience.org}), a service of LIGO Laboratory, the LIGO Scientific Collaboration and the Virgo Collaboration. LIGO is funded by the U.S. National Science Foundation. Virgo is funded by the French Centre National de Recherche Scientifique (CNRS), the Italian Istituto Nazionale della Fisica Nucleare (INFN) and the Dutch Nikhef, with contributions by Polish and Hungarian institutes.}

\bibliographystyle{aasjournal}
\bibliography{references}
\end{document}